\documentclass[preprints,conferenceproceedings,accept,moreauthors,pdftex,10pt,a4paper]{Definitions/mdpi} 
\firstpage{1} 
\pubvolume{xx}
\issuenum{1}
\articlenumber{5}
\pubyear{2018}
\copyrightyear{2018}
\history{}

\pdfoutput=1

\Title{Anisotropic flow in the few collisions regime: application to bottomonia$^{\dagger}$}

\Author{Nina Kersting*, Nicolas Borghini\orcidA{} and Steffen Feld}

\AuthorNames{Nina Kersting, Nicolas Borghini and Steffen Feld}

\address[1]{Fakult\"at f\"ur Physik, Universit\"at Bielefeld, Postfach 100131, D-33501 Bielefeld, Germany}

\corres{Correspondence: nkersting@uni-bielefeld.de}

\firstnote{Presented at Hot Quarks 2018 - Workshop for young scientists on the physics of ultrarelativistic nucleus-nucleus collisions, Texel, The Netherlands, September 7-14 2018}

\abstract{Considering the kinetic Boltzmann equation in the limit of very few collisions, we study the evolution of the phase space distribution of bottomonia interacting with an expanding gas of massless partons. We investigate the scaling of the anisotropic flow coefficients on the initial eccentricities and the inverse Knudsen number, and compute their transverse momentum dependence.}


\begin{document}

\section{Introduction}

One of the main pillars underlying the physical interpretation of experimental data from ultrarelativistic collisions of heavy nuclei at RHIC or LHC is the collective flow of outgoing particles~\cite{Heinz:2013th}. 
In particular, the measured values of the Fourier coefficients
\begin{equation}
\label{vn}
v_n(p_T) = \int\!{\rm e}^{{\rm i} n \varphi}\frac{d^2 N}{d^2 \mathbf{p}_T}\,d\varphi
  \bigg/\! \int\!\frac{d^2 N}{d^2 \mathbf{p}_T}\,d\varphi,
\end{equation}
which quantify the azimuthal anisotropy of the transverse emission pattern, can be adequately reproduced in fluid-dynamical descriptions of the created fireball expansion, where $\varphi$ denotes the azimuth of the transverse momentum $\mathbf{p}_T$ of a particle and $d^2 N/d^2 \mathbf{p}_T$ the final-state $\mathbf{p}_T$-distribution. 
The anisotropic flow coefficients $v_n$ reflect the asymmetry of the initial transverse geometry of the overlap region of the colliding nuclei, quantified by eccentricities $\epsilon_m$ and symmetry planes $\Phi_m$.

The fluid-dynamical interpretation has been challenged by measurements of very similar coefficients $v_n$ in ``smaller systems'', as created in $p+A$ or even high multiplicity $p+p$ collisions, in which the creation of a collectively behaving medium was not expected. 
This has led to a renaissance of studies utilizing Boltzmann kinetic theory in the few collisions regime of small inverse Knudsen number $Kn^{-1}$~\cite{Heiselberg:1998es,Borghini:2010hy,Romatschke:2018wgi,Kurkela:2018ygx,Borghini:2018xum}.
In such approaches, the transverse momentum distribution entering eq.~\eqref{vn} is actually the late time limit of a time-dependent distribution, which is itself related by a straightforward integral
\begin{equation}
\label{eq2}
\frac{d^2 N}{d^2 \mathbf{p_T}}(t, \mathbf{p_T}) =
\!\int\! f(t, \mathbf{x}, \mathbf{p_T})\,d^2 \mathbf{x}
\end{equation}
to the central object of the kinetic theoretical framework, viz., the phase-space distribution $f(t, \mathbf{x}, \mathbf{p_T})$, whose evolution is governed by the Boltzmann equation. 
Note that eq.~\eqref{eq2} implicitly means that we only consider a two-dimensional setup hereafter---the generalization to a three-dimensional evolution with longitudinal boost invariance, as approximately holds around midrapidity, is quite straightforward. 

While the kinetic-theoretical ansatz should certainly be relevant for the description of small and dilute systems, it may also hold for the larger ones created in $A+A$ collisions, in the case of particles that barely interact with the rest of the fireball. 
Prominent examples are high-momentum partons that can quickly escape the system~\cite{Romatschke:2018wgi}, or color-neutral objects like heavy quarkonia generated in the initial stage provided the medium energy density is such that they remain bound. 
In these proceedings we briefly present the main elements of the description and apply it to an expanding system of massless ``medium particles'' (partons, labeled with a subscript p) and few massive particles (bottomonia, labeled with B), and demonstrate that even very few collisions in average per bottomonium lead to sizable anisotropic flow coefficients.

\section{Dynamics in the few collisions regime}

\subsection{Initial condition and evolution equation}

The starting point of the model is an initial condition at time $t=0$ for each single-particle phase space distribution: for particles of type $j$ (with $j = {\rm B}$ or p), we assume
\begin{equation}
\label{initialdistributiondetailed}
f^{(0)}_j(0, \mathbf{x}, \mathbf{p}_j) = 
  \frac{N_{j\,}F(p_j)}{2\pi R_j^2} {\rm e}^{-r^2/2 R_j^2\!}
  \left(1 - 4_{}\epsilon_{2,j\,} {\rm e}^{-r^2/2 R_j^2\!} 
    \bigg(\frac{r}{R_j}\bigg)^{\!\!2}\cos\big[2(\theta - \Phi_{2,j})\big] \right) 
\end{equation}
where $(r,\theta)$ are centered polar coordinates in position space, while $R_j$ is a typical length scale of the initial overlap region.
The exact form of the initial momentum distribution $F(p_j)$ plays no role in the following, apart from its dependence on the modulus $p_j$ of the transverse momentum only, meaning that there is no anisotropic flow in the initial state, and its independence from the transverse position. 
$N_j$ is the initial number of particles of type $j$, while $\epsilon_{2,j}$ is the eccentricity of the almond-shaped spatial distribution, with a shorter axis along $\Phi_{2,j}$. 
Higher order initial eccentricities (triangularity $\epsilon_{3,j}$\dots) are easily accommodated in the model, but are left aside for brevity. 
For the curves shown in section~\ref{sec:3}, we assumed $\Phi_{2,\rm B} = \Phi_{2,\rm p}$ and $R_{\rm B} = R_{\rm p}$. 
Eventually, the function $\exp(-r^2/2 R_j^2)$ multiplying the $\cos(2\theta)$ term is a convenient cutoff function regulating the growth of $r^2$ at large $r$; other choices for that function are possible, without any significant influence on the final results. 

The ensuing evolution of the phase space distributions is governed by the relativistic Boltzmann equation
\begin{equation}
\label{Boltzmann}
p_j^\mu \partial_\mu f_j(t, \mathbf{x}, \mathbf{p}_j) = 
  C_{j,\rm coll.\!}\left[f_{\rm B},f_{\rm p}\right],
\end{equation}
where $C_{j,\rm coll.}$ denotes the collision term modeling the influence of scatterings on $f_j$.
Combining this evolution equation (integrated over $\mathbf{x}$) and eqs.~\eqref{vn} and~\eqref{eq2}, one can compute the time-dependence of the anisotropic flow coefficients $v_n$, and in particular their late-time value.

\subsection{Free streaming}

In the absence of inter-particle scatterings, which corresponds to a vanishing collision term in eq.~\eqref{Boltzmann}, i.e.\ to the homogeneous equation, the Boltzmann equation is solved by the free-streaming solution $f^{(0)}_j(t, \mathbf{x}, \mathbf{p}_j)$ that simply propagates the initial distribution at time $t=0$ according to 
\begin{equation}
\label{freestreaming}
f^{(0)}_j(t, \mathbf{x}, \mathbf{p}_j ) =
  f^{(0)}_j(0, \mathbf{x}-t \mathbf{v}_j, \mathbf{p}_j ),
\end{equation}
where $\mathbf{v}_j$ denotes the velocity corresponding to the momentum $\mathbf{p}_j$. 
As is well known, in that case the momentum anisotropy at any time is that in the initial state, i.e.\ no anisotropic flow is generated.

\subsection{Few collisions regime}

Let us now assume that the particles of each species undergo a small number of collisions per particle, such that the resulting phase space densities slightly deviate from the free-streaming ones~\cite{Borghini:2010hy}: 
\begin{equation}
f_j(t, \mathbf{x}, \mathbf{p}_j ) =
  f^{(0)}_j(t, \mathbf{x}, \mathbf{p}_j ) +  f^{(1)}_j(t, \mathbf{x}, \mathbf{p}_j ),
\end{equation}
where $f^{(1)}$, which vanishes at time $t=0$, is to be much smaller than $f^{(0)}$.
We further assume that the bottomonia are so few in the system that they do not scatter with themselves, but only on the massless partners, in such a way that such encounters always lead to their dissociation. 
That is, the final state bottomonia are those that escaped the fireball, and their flow thus results from an anisotropic escape mechanism~\cite{Bhaduri:2018iwr}.
Disregarding the possible reformation of bottomonia, we thus consider as collision term in their evolution equation
\begin{equation}
\label{Ccoll}
C_{\rm B,coll.}\left[f_{\rm B},f_{\rm p}\right] = 
  -\!\int\! f_{\rm B}(t, \mathbf{x}, \mathbf{p}_{\rm B}) 
    f_{\rm p}(t, \mathbf{x}, \mathbf{p}_{\rm p})_{}
 \sigma_{\rm Bp\,} v_{\rm Bp}\,\frac{d^2\mathbf{p}_{\rm p}}{(2\pi)^2}
\end{equation}
with $v_{\rm Bp}$ the relative velocity between the two colliding particles and $\sigma_{\rm Bp}$ the interaction cross section. 
The latter, which in our two-dimensional setup has the dimension of a length, sets the scale of the average number of collisions per bottomonium---which has to be smaller than~1 since every collision destroys the participating bottomonium!
The inverse Knudsen number $Kn^{-1}$ is thus proportional to $N_{\rm p}\sigma_{\rm Bp}/R$.
Note that Figure~\ref{plotv2andv4} was obtained with a constant cross section, which is not realistic---one would wish $\sigma_{\rm Bp}$ to be related to the medium energy density, to model the temperature dependence of bottomonium suppression---but can be quite easily improved to come closer to more elaborate predictions for bottomonium flow~\cite{Du:2017qkv}.

Since the correction terms $f_j^{(1)}$ is much smaller than the corresponding $f_j^{(0)}$, one may approximate the collision term by replacing each $f$ by the free-streaming solutions \eqref{freestreaming}, so that everything is determined by the initial state distributions~\eqref{initialdistributiondetailed}.

\section{Results and discussion -- Bottomonium anisotropic flow in the few collisions regime}
\label{sec:3}

Starting from the initial distributions~\eqref{initialdistributiondetailed}, we can compute the anisotropic flow coefficients $v_2$ and $v_4$ of bottomonia for the case of a collision kernel~\eqref{Ccoll} estimated with the free-streaming distributions. 
At a given transverse momentum (or, strictly speaking, transverse velocity, since this is the relevant kinematic quantity in the Boltzmann equation), we find
\begin{equation}
v_2\propto\frac{N_{\rm p}\sigma_{\rm Bp\,}}{R}(\epsilon_{2,\rm p} + \epsilon_{2,\rm B})
\quad,\quad
v_4\propto\frac{N_{\rm p}\sigma_{\rm Bp\,}}{R}\epsilon_{2,\rm p}\epsilon_{2,\rm B},
\end{equation}
where we left aside unimportant numerical factors.
We thus find that, to the considered level of approximation, both $v_2$ and $v_4$ are proportional to $Kn^{-1}$, as reported in Ref.~\cite{Borghini:2018xum}. 
By keeping distinct eccentricities for the initial distributions of bottomonia and massless partners, we exhibit specific scalings with $\epsilon_{2,\rm p}$ and $\epsilon_{2,\rm B}$, which to our knowledge were never noted before: 
both contribute independently to $v_2$, which is linear in the eccentricities, and together to $v_4$, which is of quadratic order. 
Note that by deliberately omitting any ``quadrangularity'' $\epsilon_4$ in the initial state, we miss the corresponding contributions to $v_4$~\cite{Borghini:2018xum}.

\begin{figure}[h]
	\centering
	\includegraphics[width=.53\linewidth]{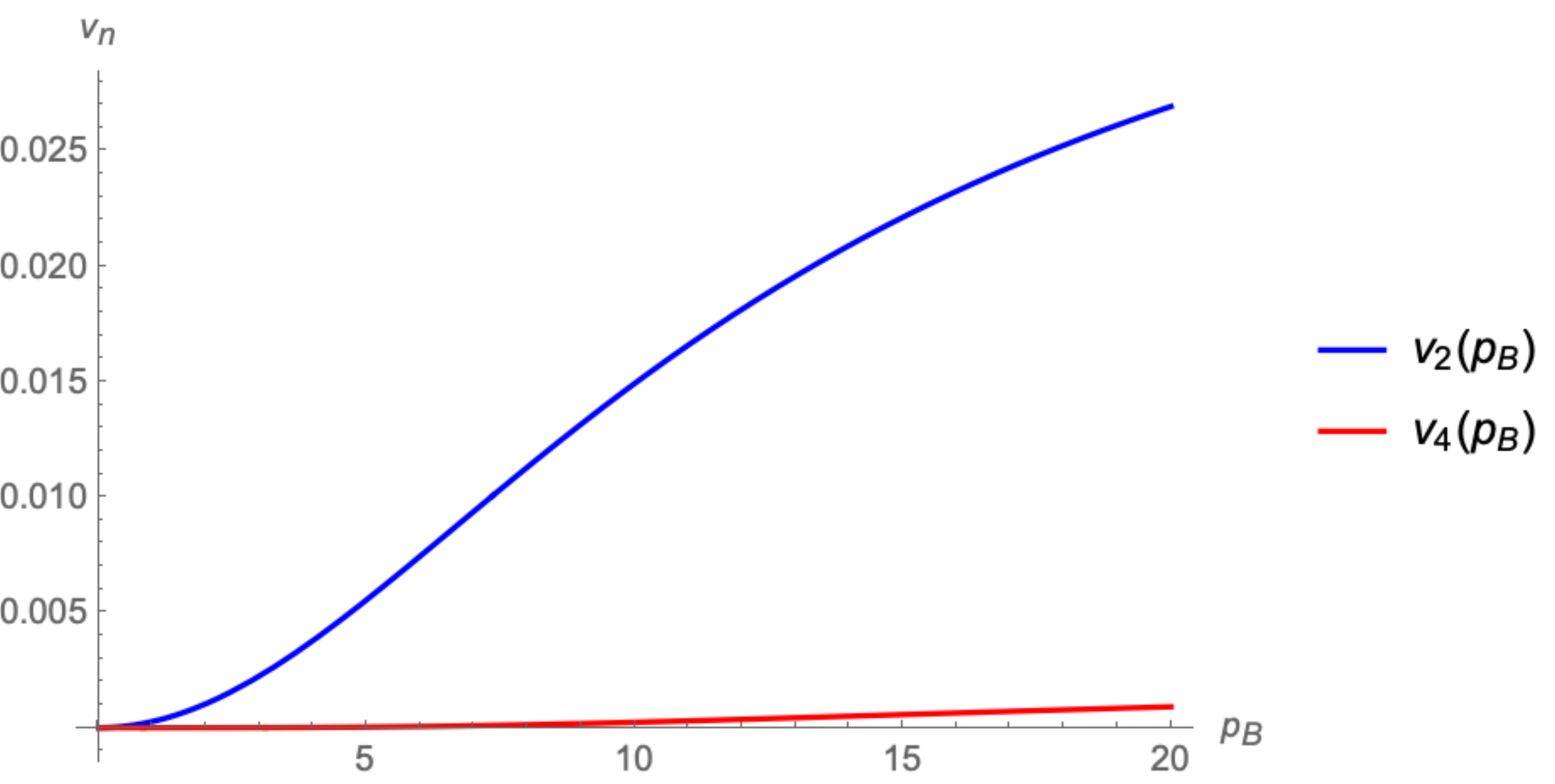}
	\caption{Elliptic flow $v_2$ and quadrangular flow 
		$v_4$ of bottomonia as a function of transverse momentum for initial eccentricities $\epsilon_2$ equal to 0.5.}
	\label{plotv2andv4}
\end{figure}   

Setting now the average number of collisions per bottomonium to 1---which is physically absurd since none is surviving, but gives an upper limit on the signal---and taking both eccentricities to be equal to 0.5, we show in Figure~\ref{plotv2andv4} the transverse momentum dependence of these anisotropic flow coefficients. 
In comparison to more elaborate computations~\cite{Du:2017qkv,Bhaduri:2018iwr}, both $v_2$ and $v_4$ grow over the whole momentum range: 
in contrast to those models, we have no mechanism to ``protect'' the bottomonia from destruction, like the competition between their formation time---there is none in our model---and the time scale for the drop of the disassociation cross section, which we take to be constant. 

Nonetheless the main result from our approach should be clear, namely that the considered mechanism efficiently produces sizeable anisotropic flow coefficients $v_n$---although $v_4$ is at the limit of what is measurable.
We admittedly enhanced the signals by going to the maximal acceptable value for $Kn^{-1}$ and implementing a constant cross section, which means that the bottomonia feel the asymmetric shape---which naturally decreases with time---of the fireball along the whole evolution. 
On the other hand, by considering a free-streaming medium of massless constituents, we maximize its expansion rate and thus its rarefaction, which in turn works in the opposite direction, diminishing the signal. 

An important feature of our approach is that we can improve the model in a rather systematic way, while still remaining at the semi-analytical level for the calculation of the anisotropic flow coefficients $v_n$, at least as long as one keeps the assumption of a small number $Kn^{-1}$, which is anyway meaningful for bottomonia. 
Of course, a natural extension is to add the longitudinal space dimension.
We can further distort the initial distributions, either including higher-order eccentricities, which will give rise to other anisotropic flow harmonics, or examining how differences between the distributions (symmetry planes, sizes) of the two particle species influence the final results.  
Another direction is to modify the ansatz for the distribution of partons directly in the collision integral, going for instance to an equilibrium distribution corresponding to a thermalized medium. 
As was mentioned above, we will also implement more realistic cross sections. 

All in all, we can explore which ingredients allow for a better comparison to experimental data, at least for heavy quarkonia, for which the model should be helpful both in $A+A$ collisions and in smaller systems.
Our present and subsequent findings need also be compared with numerical studies done in this direction, to support their calibration in the few collisions regime.



\vspace{6pt} 




\funding{We thank support by the Deutsche Forschungsgemeinschaft (DFG) through the grant CRC-TR 211 “Strong-interaction matter under extreme conditions”.}






\reftitle{References}




\end{document}